# Nonlinear DNA dynamics: nonlinearity versus dispersion

Slobodan Zdravković[a,*], Miljko V. Satarić[b]

[a] *Institut za nuklearne nauke Vinča, Univerzitet u Beogradu, Poštanski fah 522, 11001 Beograd, Serbia*
[b] *Fakultet tehničkih nauka, Univerzitet u Novom Sadu, 21000 Novi Sad, Serbia*

*Keywords:*
Nonlinear Schrödinger equation
Nonlinearity
Dispersion
DNA dynamics
Helicoidal Peyrard-Bishop model

A B S T R A C T

In the present paper we study the impact of dispersion and nonlinearity on DNA dynamics. We rely on the helicoidal Peyrard-Bishop model and use the fact that nonlinear DNA dynamics represents an interplay between nonlinearity and dispersion. We state that a dispersion coefficient $P$ and a coefficient of nonlinearity $Q$, existing in nonlinear Schrödinger equation, are mutually dependent and show how function $Q(P)$ and $P(Q)$ can be obtained. Also, we show how all this can be used to find a possible interval for the parameter describing helicoidal structure of DNA.

## 1. Introduction

Nonlinear Schrödinger equation (NLSE) has been used in many branches of physics and mathematics. One of its solutions, having profound physical meaning, is a modulated solitonic wave. An example, relevant for this paper, where NLSE is applied, is nonlinear dynamics of DNA. To study this dynamics we rely on helicoidal Peyrard-Bishop (HPB) model [1].

Nonlinear dynamics of DNA can be viewed as an interplay between nonlinearity and dispersion [2]. This means that nonlinearity increases wave amplitude and decreases its width, while the impact of dispersion on the wave is opposite. This is demonstrated in this paper. In particular, we show that the mentioned relationships between the wave characteristics and nonlinearity and dispersion hold if nonlinear and dispersion parameters, existing in NLSE, are mutually dependent. We study how they depend on each other and show that this procedure can be used to determine possible intervals of some internal parameters, describing chemical interactions and geometry of DNA. The method explained here can be used in all branches of science where NLSE is used and some parameters are not known.

The paper is organized as follows. In Section 2 we give a brief review of HPB model of DNA. In the next two sections we study the relationships between nonlinearity and dispersion while Section 5 comprises some concluding remarks.

## 2. A brief review of the HPB model

---

[*] Corresponding author.
 *E-mail address:* szdjidji@vinca.rs.

It is well known that DNA molecule is a double helix and each strand is a polymeric collection of nucleotides. The nucleotides belonging to the same strand are connected by strong covalent bonds, modelled by the nearest-neighbour harmonic interactions, while the strands are coupled to each other through the weak hydrogen bonds, modeled by Morse potential. In what follows we very briefly describe the HPB model. All important details and derivations can be found in a review paper [3].

The HPB model takes only transversal displacements of nucleotides into consideration. Helicoidal structure of DNA chain is taken into account assuming that a nucleotide at the site $n$ of one strand interacts with both ($n+h$)th and ($n-h$)th nucleotides of the other strand [1,3]. We here use $h=5$ [3].

Let $u_n$ and $v_n$ be the displacements of the nucleotides belonging to different strands at the position $n$ from their equilibrium positions along the direction of the hydrogen bond. If $\dot{u}_n$ and $\dot{v}_n$ represent the pertaining velocities then the Hamiltonian, describing DNA dynamics, is

$$H = \sum \left\{ \frac{m}{2}(\dot{u}_n^2 + \dot{v}_n^2) + \frac{k}{2}[(u_n - u_{n-1})^2 + (v_n - v_{n-1})^2] \right.$$
$$\left. + \frac{K}{2}[(u_n - v_{n+h})^2 + (u_n - v_{n-h})^2] + D[e^{-a(u_n - v_n)} - 1]^2 \right\}, \quad (1)$$

where $m$ is a common mass for all the nucleotides and the parameters $k$ and $K$ are the harmonic constants of the longitudinal and helicoidal springs, respectively. The last term is the Morse potential where $D$ and $a$ are the depth and the inverse width of the potential well, respectively. The Hamiltonian (1) brings about the following nonlinear dynamical equation of motion

$$m\ddot{y}_n = k(y_{n+1} + y_{n-1} - 2y_n) - K(y_{n+h} + y_{n-h} + 2y_n) + 2\sqrt{2}aD(e^{-a\sqrt{2}y_n} - 1) e^{-a\sqrt{2}y_n}, \quad (2)$$

where a new coordinate $y_n = (u_n - v_n)/\sqrt{2}$, describing the out-of-phase transversal motion, represents a stretching of the nucleotide pair at the position $n$ [1,3]. To solve Eq. (2) we assume that the nucleotides oscillate around the bottom of the Morse potential well, i.e.

$$y_n = \varepsilon \Phi_n; \qquad (\varepsilon \ll 1) \quad (3)$$

and look for wave solutions of the form

$$\Phi_n(t) = F_1(\xi)e^{i\theta_n} + \varepsilon[F_0(\xi) + F_2(\xi)e^{i2\theta_n}] + \text{cc} + O(\varepsilon^2) \quad (4)$$

$$\xi = (\varepsilon nl, \varepsilon t), \qquad \theta_n = nql - \omega t, \quad (5)$$

where $l$ is the distance between two neighbouring nucleotides in the same strand, $\omega$ is the optical frequency of the linear approximation, $q$ is the wave number whose role will be discussed later, cc represents complex conjugate terms and the function $F_0$ is real. Notice four parameters in Eq. (1) and one more in Eq. (5). These are: $k$, $K$, $D$, $a$ and $q$.

One can show that the functions $F_0$ and $F_2$ can be expressed through $F_1$, which is a solution of the well-known NLSE

$$iF_{1\tau} + PF_{1SS} + Q|F_1|^2 F_1 = 0, \quad (6)$$

where the dispersion coefficient $P$ and the coefficient of nonlinearity $Q$ are given by

$$P = \frac{1}{2\omega}\left\{\frac{l^2}{m}[k\cos(ql) - Kh^2\cos(qhl)] - V_g^2\right\} \tag{7}$$

and

$$Q = -\frac{2a^2 D}{m\omega}[2\alpha(\mu + \delta) + 3\beta]. \tag{8}$$

To derive Eq. (6) a continuum limit $nl \to z$ and new coordinates $\tau = \varepsilon^2 t$ and $S = \varepsilon(z - V_g t)$ were introduced [1,3]. The relevant terms involved in the expressions for $P$ and $Q$ are [3]:

$$\omega^2 = (4/m)\left[a^2 D + k\sin^2(ql/2) + K\cos^2(qhl/2)\right], \tag{9}$$

$$V_g = \frac{l}{m\omega}[k\sin(ql) - Kh\sin(qhl)], \tag{10}$$

$$\alpha = \frac{-3a}{\sqrt{2}}, \quad \beta = \frac{7a^2}{3}, \quad \mu = -2\alpha\left(1 + \frac{K}{a^2 D}\right)^{-1} \tag{11}$$

and

$$\delta = a^2 D\alpha\left[m\omega^2 - k\sin^2(ql) - K\cos^2(qhl) - a^2 D\right]^{-1}. \tag{12}$$

It was suggested that the corresponding wave length covers an integer number of nucleotides, i.e. [4]

$$q = \frac{2\pi}{\lambda}, \quad \lambda = Nl, \quad N \text{ integer}. \tag{13}$$

Therefore, one can assume $N$ as the internal parameter instead of $q$. The interval

$$7 \leq N \leq 20, \tag{14}$$

was suggested [5], corresponding to $0.09\,\text{Å}^{-1} \leq q \leq 0.26\,\text{Å}^{-1}$, as the well known value for the distance is $l = 0.34\,\text{nm}$.

The final expression for stretching of the nucleotide pair at the position $n$ is

$$y_n(t) = 2A\,\text{sech}\left(\frac{nl - V_e t}{L}\right)\left\{\cos(\Theta nl - \Omega t) + A\,\text{sech}\left(\frac{nl - V_e t}{L}\right)\left[\frac{\mu}{2} + \delta\cos(2(\Theta nl - \Omega t))\right]\right\}, \tag{15}$$

representing the modulated soliton. As was mentioned above all the formulas and detailed derivations can be found in Ref. [3]. It suffices now to show how its amplitude $A$ and the wave width $L$ depend on $P$ and $Q$. These crucial relationships are:

$$A \propto \frac{1}{\sqrt{PQ}} \equiv \frac{1}{\sqrt{f(P,Q)}}, \qquad L \propto P, \tag{16}$$

where $P > 0$ and $Q > 0$ [3,6], while $f$ stands for a function of $P$ and $Q$.

It was stated above that nonlinear dynamics of DNA should be viewed as an interplay between nonlinearity and dispersion. This means that $A$ and $L$ should depend on $Q$ and $P$ in the following way:

$$A(P)\downarrow, \quad A(Q)\uparrow, \quad L(P)\uparrow, \quad L(Q)\downarrow, \tag{17}$$

where the arrows $\uparrow$ and $\downarrow$ mean increasing and decreasing functions, respectively. In the next section we will show that $P$ and $Q$ are mutually dependent and will obtain functions $Q(P)$ and $P(Q)$, satisfying Eq. (17).

## 3. Nonlinearity vs. dispersion

If we look at Eqs. (16) we do not see that Eq. (17) is satisfied. One might think that $A$ equaly depends on both parameters and that the wave width does not depend on $Q$ at all. Of course, this is wrong and certainly indicates that $P$ and $Q$ are mutually dependent. For example, if $P \propto Q^n$ then Eq. (17) holds for $n < -1$. Hence, our next task is to obtain a function $Q(P)$.

Let us remind ourselves that $P$ and $Q$ depend on five internal parameters. These are: $k$, $K$, $D$, $a$ and $N$. The values of these parameters are still not available from experiments or ab initio calculations. We have done a big effort trying to estimate them [5,7-10]. For example, we estimated the window of a ratio $k/K$ [9,10]. The approximate interval would be

$$70 < k/K < 210. \tag{18}$$

Of course, each interval depends on the assumed values for the remaining parameters.

The most intriguing of all the internal parameters is $K$, which reflects the helicoidal structure of DNA chain. If we pick up particular values for the remaining four parameters we obtain the functions $P(K)$ and $Q(K)$. From them, we can obtain the relationship between $P$ and $Q$. The function $Q(P)$ is shown in Fig. 1 for two chosen values of $N$ and for a certain combination of the remaining three parameters [5]. For the parameter $K$ we picked up the interval

$$0.05\,\text{N/m} < K < 0.15\,\text{N/m}, \tag{19}$$

which approximately matches the interval (18), for the chosen value of the parameter $k$. The figure obviously indicates a linear decreasing function. In what follows we study whether Eq. (17) is satisfied for

$$Q = Q_0 - C_0 P, \tag{20}$$

where $Q_0$ and $C_0$ are positive constants. The second step will be $Q$ as a polynomial of $P$. All the values for $P$ and $Q$ are expressed in units

$$P_{\mathrm{u}} = 10^{-8}\,\mathrm{m^2/s} = \mathring{\mathrm{A}}^2/\mathrm{ps}, \qquad Q_{\mathrm{u}} = 10^{32}\,1/\mathrm{m^2 s} = 1/\mathring{\mathrm{A}}^2\,\mathrm{ps}. \tag{21}$$

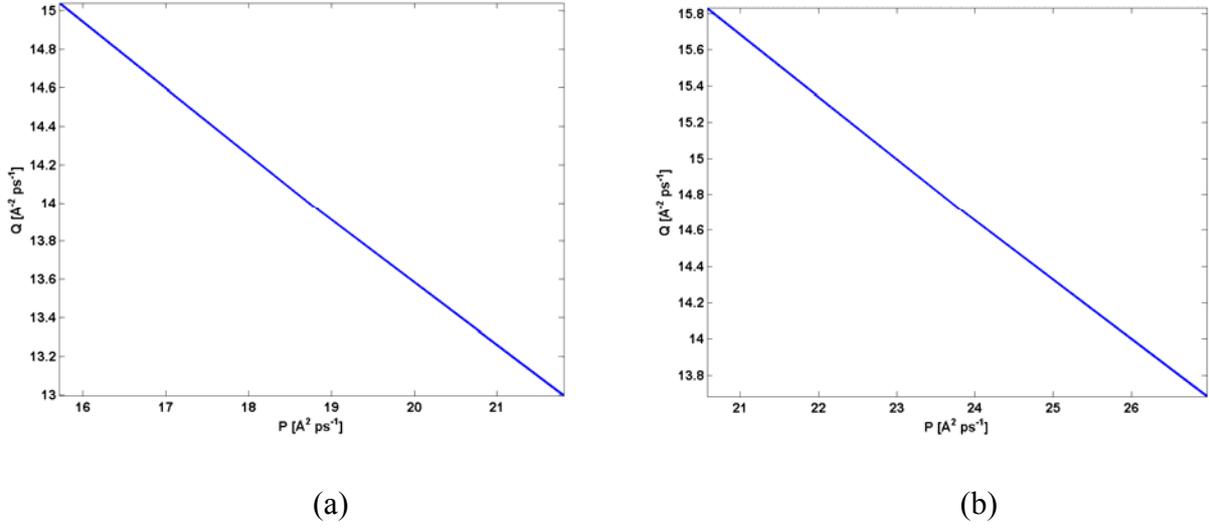

(a)          (b)

**Fig. 1.** A function $Q(P)$ for: $a = 1.2\,\mathring{\mathrm{A}}^{-1}$, $D = 0.07\,\mathrm{eV}$, $k = 12\,\mathrm{N/m}$ and: (a) $N = 10$; (b) $N = 12$.

One can easily show that the function $f$, given by Eq. (16), is an increasing one for

$$0 < P < \frac{Q_0}{2C_0}, \tag{22}$$

which is a requirement for the first of Eqs. (17) to be satisfied. Also, the second one holds for

$$\frac{Q_0}{2} < Q < Q_0, \tag{23}$$

while the remaining two are satisfied without any additional requirement.

Therefore, we now know the intervals for $P$ and $Q$ that have physical meaning. As the functions $P(K)$ and $Q(K)$ are known we can, using a computer, easily find the appropriate interval for $K$. This will be demonstrated for $N = 10$. Notice that for this particular value for $N$ the function $Q(P)$ can be obtained analytically as the function $K(P)$ exists. Let us pick up two points for the linear function, determined by $K_1 = 0.07\,\mathrm{N/m}$ and $K_2 = 0.13\,\mathrm{N/m}$. According to the functions $P(K)$ and $Q(K)$ we can obtain the appropriate values for $P_i$ and $Q_i$ ($i = 1, 2$), which yields

$$C_0 = 0.336, \qquad Q_0 = 20.30. \tag{24}$$

Of course, all the units are determined by Eq. (21). Finally, using the computer and the functions $P(K)$ and $Q(K)$ we can easily show that Eqs. (22) and (23) give the intervals $K < 0.29\,\mathrm{N/m}$ and $0 < K < 0.31\,\mathrm{N/m}$, respectively. Therefore, the conclusion is

$$K < 0.29\,\mathrm{N/m}, \tag{25}$$

which is much broader interval than the assumed one, given by Eq. (19). This means that our initial choice was good. This is so because Eq. (19) is a result of the previous investigations. However, the values of the parameters are usually not known. Without precise experimental results, the best that the theoreticians can do is to shorten the allowed intervals as much as possible. For this purpose we should combine all the known procedures. The one used in this paper is simple and may be convenient to start with. Of course, when we obtain a few intervals corresponding to different methods we accept the overlapping values. We should keep in mind that NLSE is widely used. Very often the researchers have almost no idea about the possible values of a certain parameter. If so then we should start with much wider interval. In the next section we study a general approach. We start with a large interval for $K$ and look for $Q(P)$ as a polynomial.

## 4. General approach

A purpose of this section is to increase precision of the procedure explained in the previous one. To achieve this goal, the functions $P(Q)$ and $Q(P)$ are expressed as the third order polynomials.

Let us start with

$$K < 0.40 \, \text{N/m}, \tag{26}$$

which is much wider interval than the one in Eq. (19). The function $P(Q)$ can be obtained using a computer. It is shown in Fig. 2 together with its fit. It is not linear any more and should be represented by the polynomial. The third order polynomial is

$$P = -2.240 \times 10^{-3} Q^3 + 0.1725 Q^2 - 6.492 Q + 81.95. \tag{27}$$

The same procedure can be done for $Q(P)$ and the corresponding polynomial is

$$Q = -2.247 \times 10^{-5} P^3 + 4.206 \times 10^{-3} P^2 - 0.4695 P + 21.47. \tag{28}$$

We do not have to use higher order polynomials as the precision of the third order polynomial is $10^{-4}$. Of course, both $P$ and $Q$ are expressed in units given by Eq. (21).

According to Eqs. (16) and (27) we can easily express the amplitude $A$ as a function of $Q$. This is a decreasing function for $Q > 9.083$. This cut off gives the highest allowed value for the parameter $K$. Namely, according to Eqs. (8), (9), (11)-(13) we easily obtain the function $Q(K)$, which is

$$Q(K) = -\frac{13.71 \, (-8.736 + K)(-1.148 + K)}{(-5.276 + K)(1.613 + K)}. \tag{29}$$

This function is shown in Fig. 3. Hence, $Q > 9.083 \times 10^{32} \, \text{m}^{-2}\text{s}^{-1}$ for

$$K < 0.375 \, \text{N/m}. \tag{30}$$

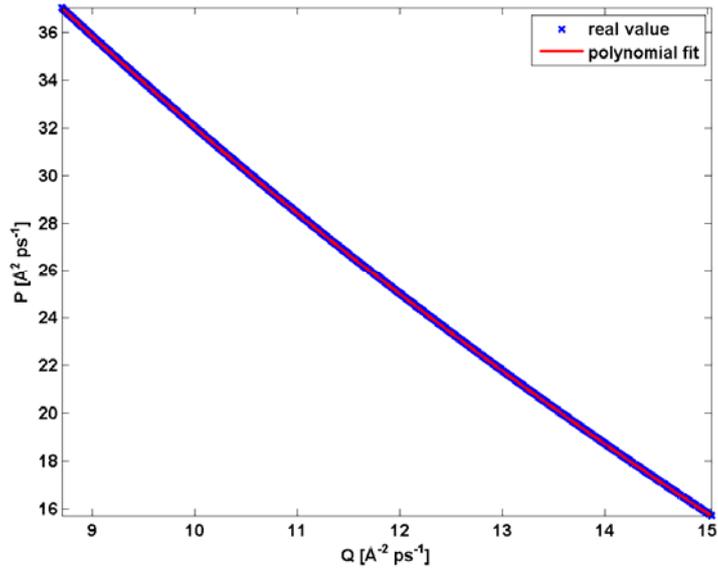

**Fig. 2.** A function $P(Q)$ as a third order polynomial.

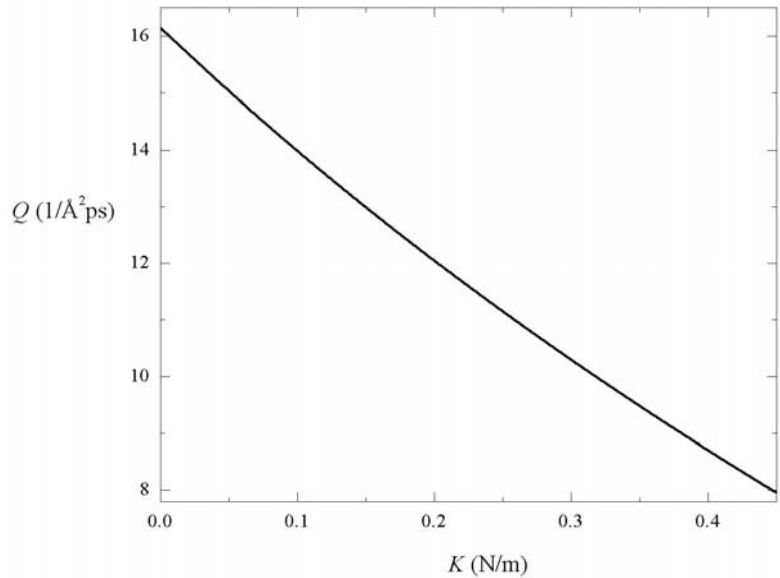

**Fig. 3.** A function $Q(K)$ for: $a = 1.2\,\text{Å}^{-1}$, $D = 0.07\,\text{eV}$, $k = 12\,\text{N/m}$ and $N = 10$.

If we used Eq. (28) instead of Eq. (27) we would obtain the function $A(P)$. This is the increasing function for $P < 35.52 \times 10^{-8}\,\text{m}^2/\text{s}$. The curve $P(K)$ brings about Eq. (30) again.

It is obvious that the value given by Eq. (25) is more precise than the one given by Eq. (30). This does not mean that the linear function ensures better results than the polynomials because the initial interval for $K$ was bigger in the latter case.

## 5. Concluding remarks

In this paper we demonstrated the expected interplay between nonlinearity and dispersion regarding the solution of NLSE. Also, we showed how the impact of nonlinear and dispersion parameters on the amplitude and the wave length of the solitonic wave can be used to study the possible values of unknown intrinsic parameters.

Nonlinear DNA dynamics was picked up as an particularly interesting example. However, this procedure is rather general as NLSE has been applied in many branches of physics and mathematics [11]. Some examples could be plasma physics [12,13] and nonlinear optics [14,15]. This famous equation has also been used to study oceanic waves [16,17], nonlinear electrical lines [6], Bose-Einstein condensates [18] and so on. Therefore, our procedure is rather general and can be applied wherever NLSE is used and the values of some parameters are not known.

**Acknowledgments**

We thank Aleksandra Maluckov for bringing a couple of references to our attention and to Jovana Petrović who plot Fig. 2 for us.

This research was supported by funds from Serbian Ministry of Education and Sciences, grants III45010 and OI171009.